\documentclass[a4paper,11pt]{article}
\usepackage{pos}
\usepackage{braket}
\usepackage{tikz}
\usepackage{wrapfig}
\usepackage{orcidlink}
\usetikzlibrary{arrows.meta}
\usetikzlibrary{positioning}
\usetikzlibrary{decorations}
\usetikzlibrary{decorations.markings}

\usepackage{graphicx} 
\usepackage{float}

\DeclareMathOperator{\SU}{\mathrm{SU}}

\newcommand{\eq}{\begin{equation}}   
\newcommand{\en}{\end{equation}}   
\newcommand{\eqa}{\begin{eqnarray}}   
\newcommand{\ena}{\end{eqnarray}}   

\newcommand{\bea}{\begin{eqnarray}}   
\newcommand{\eea}{\end{eqnarray}}

\title{Lattice Determination of the Baryon Junction Mass in $(2+1)$ Dimensions}

\author*[a]{Dario Panfalone}
\author[a]{Michele Caselle}
\author[b]{Nicodemo Magnoli}
\author[a]{Lorenzo Verzichelli}

\affiliation[a]{Dipartimento di Fisica,  Universit\'a degli Studi di Torino and INFN, Sezione di Torino, \\
Via Pietro Giuria 1, I-10125 Turin, Italy}

\affiliation[b]{Department of Physics, University of Genoa and INFN, Sezione di Genova, \\
Via Dodecaneso 33, I-16146, Genoa, Italy}

\emailAdd{dario.panfalone@unito.it}
\emailAdd{caselle@to.infn.it}
\emailAdd{magnoli@ge.infn.it}
\emailAdd{lorenzo.verzichelli@unito.it}

\abstract{This contribution investigates baryonic flux tube configurations in $\SU(3)$ Yang--Mills theory in $(2+1)$ dimensions. Leveraging recent next-to-leading-order results within the Effective String Theory (EST) framework, which explicitly include corrections proportional to the baryon junction mass $M$ up to order $1/R^2$, we carry out a non-perturbative determination of this parameter, through high-precision simulations of the three-point Polyakov-loop in the open string channel. In addition, the high-temperature regime of the baryonic system is examined in order to test the Svetitsky--Yaffe conjecture. Close to the deconfinement transition, the lattice results for the correlators show close agreement with the predictions of the two-dimensional three-state Potts model.}
\FullConference{The 42nd International Symposium on Lattice Field Theory (LATTICE2025)\\
2-8 November 2025\\
Tata Institute of Fundamental Research, Mumbai, India\\}

\begin{document}
\maketitle

\section{Introduction and motivation}
In Yang-Mills theory, color confinement manifests through the formation of chromoelectric flux tubes between distant color sources, resulting in a linearly increasing potential. Effective String Theory (EST) provides a robust framework for modeling this non-perturbative phenomenon describing the flux tube as a thin, fluctuating  string~\cite{Nambu:1974zg,Goto:1971ce,Luscher:1980ac,Polchinski:1991ax}.
Recently the so-called "low energy universality" was found, a crucial feature characterizing the EST. This property predicts that the leading terms in the long-distance expansion of the effective string action are not arbitrary, but universal, and correspond to the Nambu-Goto (NG) action ~\cite{Luscher:2004ib, Billo:2012da, Gliozzi:2012cx, Aharony:2013ipa, Brandt:2016xsp, Caselle:2021eir}.

However, it is now well-established that the Nambu-Goto action does not offer the complete description of the EST. 
The terms beyond the Nambu-Goto (BNG) approximation appear only at high orders in the expansion around the limit of an infinitely long string, making them difficult to estimate. Despite their subleading nature, these terms encode the information necessary for a precise characterization of the infrared dynamics of confining theories. While the leading-order universality explains the striking similarities across different gauge groups, the specific signatures that distinguish these theories are contained within the BNG corrections.

Typically, EST predictions are tested on the lattice by studying the Polyakov loop correlator, from which one can extract the potential between two color charges. 

In this context, a systematic study of the effective string corrections beyond the Nambu–Goto action was performed in Ref.~\cite{Caselle:2024zoh} for SU(N) Yang–Mills theories in 2+1 spacetime dimensions. The primary objective of that analysis was to investigate the fine details characterizing the confining dynamics across various gauge groups from high-precision Monte Carlo simulations of the two-point Polyakov loop correlation function.

A rather non-trivial test of the EST approach is the description of baryon, a configuration where, in the case of $\SU(N)$ gauge theories, $N$ confining strings meet at a common point which is usually called the \textit{Baryon Junction}. 

To this problem a lot of effort was devoted, more than twenty years ago, the dominant term of the EST correction of this configuration was obtained ~\cite{Jahn:2003uz,deForcrand:2005vv,Pfeuffer:2008mz}. From the numerical point of view the study focused to the determination of the ground state and the shape of the flux tubes in presence of a baryon junction \cite{Alexandrou:2002sn,Takahashi:2002bw,Bissey:2006bz,Cardoso:2008sb,Bakry:2014gea,Borisenko:2018zzd,Koma:2017hcm,Ma:2022vqf,Dmitrasinovic:2009ma,Huebner:2007yzb,Brambilla:2009cd,Sakumichi:2015rfa}. An open question is the determination of the baryon junction mass, a quantity both relevant from the theoretical and from the phenomelogical point of view ~\cite{Karliner:2016zzc}. In a recent next-to-leading order calculation~\cite{Komargodski:2024swh}, the correction proportional to the baryon junction mass was made evident, both in the closed and in the open string channel, opening the possibility of its non-perturbative determination. 
This contribution concerns our recent work~\cite{Caselle:2025elf}, where, for the first time, we provided a numerical estimate of the baryon junction mass for the $\SU(3)$ theory in $(2+1)$ dimensions. In addiction, we will discuss the numerical data in the high-temperature regime of the theory, providing a new impressive test of the Svetitsky–Yaffe conjecture.

\section{Baryons in EST}
In this section we briefly review the results of Ref.~\cite{Komargodski:2024swh}, where, for $\mathrm{SU}(3)$ gauge theory in $(d+1)$ dimensions, the next-to-leading-order correction associated with the term proportional to the baryon junction mass was computed, and we focus on the $(2+1)$-dimensional case, which corresponds to the setup adopted in our simulations.

The expression for the three-point Polyakov loop correlator can be expanded in the two limits of "open" and "closed" string channel.

In the open string channel ($N_t \gg R$) one obtains
\begin{equation}
\braket{P(x_1) \, P(x_2) \, P(x_3)}    = A_{open} (N_t) \, e^{-N_t E_0(R)},
\label{eqlowT}
\end{equation}
where $x_1,x_2$ and $x_3$ are the three vertices of an equilateral triangle, $R$ denotes the distance of these poits with respect to the Fermat point and $N_t$ is the size of the lattice in the compactified euclidead time direction.

The prefactor $A_{open} (N_t) = A (N_t) e^{-N_t M}$ incorporates an unknown functional form $A(N_t)$. This term exhibits a complex dependence on $N_t$ resulting from the Dirichlet boundary conditions at the vertices. In this setting, such dependence complicates the isolation and subsequent extraction of the baryon junction mass from this term.

In this set-up, the convenient way to estimate $M$ is studyng the the $R$ dependence, which is all encoded in the ground state energy, where the first correction beyond the linearly rising term is the $1/R^2$ term due to the baryon junction: 
 \begin{equation}
        E_{0}(R) =  3R\sigma - \frac{M \pi}{48 \sigma R^2} + \mathcal{O}(1/R^3).
        \label{eq2}
\end{equation}

On the other hand, in the closed string limit ($R \gg N_t$ with $N_t\sqrt{\sigma} \gg 1$), one finds
\begin{equation}
 \braket{P(x_1) P(x_2) P(x_3)}    = A_{closed}(N_t)\sqrt{\frac{N_t}{R}} e^{-3 R E_0(N_t)},
\label{eqhighT}
\end{equation}
with
 \begin{equation}
    E_0(N_t)=\sigma N_t\left(1-\frac{\pi}{6\sigma N_t^2} + \mathcal{O}\left( 1 / {N_t}^4 \right) \right)
        \label{eq5},
\end{equation}
and
\begin{equation}
A_{\text{closed}}(N_t) =
A(N_t)\,
e^{
- M N_t
\left[
1 - \frac{2\pi}{9\sigma N_t^2}
+ \mathcal{O}\!\left(1/N_t^4\right)
\right]}
\end{equation}

EST results are large distance expansions, in this case, in powers of  $\frac{1}{N_t\sqrt{\sigma}}$. In the closed-string channel, the focus is usually on the high-temperature region, close to the deconfinement transition while remaining within the confined phase. In this regime $N_t$ is small and the constraint $R \gg N_t$ can be achieved with reasonable numerical cost. However, since $N_t\sqrt{\sigma}\sim 1$, subleading contributions in the Nambu–Gotō effective action are no longer negligible. In the absence of explicit EST predictions for these higher-order terms, one can adopt an alternative strategy based on the Svetitsky–Yaffe conjecture, which allows to map, in the vicinity of the deconfinement transition, the $\SU(3)$ gauge model in (2+1) dimensions to the three-states Potts model in two dimensions. As discussed below, comparing these two approximations allows for an estimation of the higher-order Nambu--Got\=o corrections in the baryonic potential.

\subsection{Comparison with Svetitsky–Yaffe predictions}

The $R$ dependence of eq.~\eqref{eqhighT} is remarkably similar to the expression the two point correlator obtained in the same limit of closed string channel:
\begin{equation}
\label{eq:two-pt}
\braket{P^+(0)P(R)} \sim \frac{e^{-R E_{P^{\scriptscriptstyle+}P}(N_t)}}{\sqrt{R}},
\end{equation}
with 
\begin{equation}
E_{P^+P}(N_t)=\sigma N_t\sqrt{1-\frac{\pi}{3\sigma N_t^2}}.
\label{eq9}
\end{equation}
In particular, if we expand $E_{P^+P}$ at the first order (the same order considered in the baryonic case), we find extactly the same ground state energy of equation \eqref{eq5}.

This suggests to consider \eqref{eq5} only as the first order term of a large distance expansion, and that higher order corrections to the three quarks potential can be resummed exactly as follows:

\begin{equation}
    E_0(N_t)=\sigma N_t\left(1-\frac{\pi}{6\sigma N_t^2}\right) \hskip 0.3cm \longrightarrow \hskip 0.3cm
E_0(N_t)=\sigma N_t\sqrt{1-\frac{\pi}{3\sigma N_t^2}}.
\label{eq8bis}
\end{equation}

The impact of this resummation is negligible at low and intermediate temperatures. Conversely, its contribution becomes significant as the system approaches the deconfinement transition.

This suggestion is supported by the results obtained in the three-states Potts model in two dimensions~\cite{Caselle:2005sf,Guida:1995kc}, which we will briefly review in the following. Representing each spin as phase $s = e^{2 i \pi \, n / 3}$, the large distance expansion of the two-point spin correlators and of the correlator of three spins, in the case of triangles that are close to being equilateral, reads:

\begin{equation}
    \braket{s^+(0)s(R)} \ \sim \  K_0(mR)
\label{eq11},
\end{equation}
\begin{equation}
    \braket{s(x_1)s(x_2)s(x_3)} \ \sim \  K_0(m R_Y).
\end{equation}

Where $K_0(x)$ denotes the modified Bessel function of order zero, $m$ is the mass scale of the model (i.e., the inverse of the correlation length) and it is assumed that all the angles of the triangle $(x_1,x_2,x_3)$ are less than $2 \pi /3$ and   
$R_Y$ is the minimal total length of lines connecting the 3 spins to the Fermat Point.

Recalling the large distance expansion of the Bessel funcion $  K_0(x)\sim \sqrt{\frac{\pi}{2x}} \, e^{-x}$,one finds that the Potts model predictions coincide with the EST formulas in eqs.~\eqref{eq:two-pt} and~\eqref{eqhighT} once the Potts mass parameter $m$ is identified with the ground-state energy $E_0(N_t)$. Morevoer we conclude that the two and three-point functions must share the same mass scale. The same pattern emerges in the EST framework if the prediction for the two-point function is truncated at the same order adopted in the baryonic calculation. It should be emphasized, however, that the resummation in eq.~\eqref{eq8bis} is not derived within EST itself, but inferred from the Svetitsky--Yaffe mapping and from the knowledge of the two-point function. Hence, one expects corrections as the deconfinement temperature is approached, in analogy with the behavior of the ground state in the two-point correlator (see ref.~\cite{Caselle:2024zoh}). In the following, we use our Monte Carlo results to test these observations.
\section{Numerical results} 
\begin{figure}[!htb] \centering 

	\tikzset{
		lattice point/.style={
			draw,
			fill,
			circle,
			minimum size=1.0mm,
			inner sep=0,
		},
		intermediate point/.style={
			draw,
			fill,
			circle,
			minimum size=0.0mm,
			white,
			inner sep=0,
		},    
	}
	
	\begin{tikzpicture}[scale=1.1, transform shape]
		
		\draw [ -> ] (0, 0) -- (0, 1) node [pos=0.95, anchor=south] {$t$};
		\draw [ -> ] (0, 0) -- (1, 0) node [pos=0.95, anchor=north] {$x$};
		\draw [ -> ] (0, 0) -- (0.5, 0.5) node [pos=0.95, anchor=south] {$y$};
		
		\draw [blue, ultra thick, ->] (1.5, -0.5) -- (1.5, 1.6) node [anchor=east] {$P(0, 0)$};
		\draw [blue, ultra thick] (1.5, 1.3) -- (1.5, 3.5);
		
		\draw [blue, ultra thick, ->] (5.5, -0.5) -- (5.5, 1.6) node [anchor=north west] {$P(b, 0)$};
		\draw [blue, ultra thick] (5.5, 3.5) -- (5.5, 1.3);
		
		\draw [blue, ultra thick, ->] (5.0, 1.0) -- (5.0, 3.1)
		node [anchor=south east] {$P(b/2, h)$};
		\draw [blue, ultra thick] (5.0, 3.0) -- (5.0, 5.0);
		
		\draw [ultra thin] (1.5, 1.5) -- (5.5, 1.5);
		\draw [ultra thin] (2.0, 2.0) -- (6.0, 2.0);
		\draw [ultra thin] (2.5, 2.5) -- (6.5, 2.5);
		\draw [ultra thin] (3.0, 3.0) -- (7.0, 3.0);
		
		\draw [ultra thin] (1.5, 1.5) -- (3.0, 3.0);
		\draw [ultra thin] (2.5, 1.5) -- (4.0, 3.0);
		\draw [ultra thin] (3.5, 1.5) -- (5.0, 3.0);
		\draw [ultra thin] (4.5, 1.5) -- (6.0, 3.0);
		\draw [ultra thin] (5.5, 1.5) -- (7.0, 3.0);
		
		\filldraw (4.0, 2.0) circle (1 pt);
		
		
		\draw [ <-> ] (1.6, 0.5) -- (5.4, 0.5) node [pos=0.5, anchor=north] {$b$};
		\draw [ <-> ] (1.6, 1.35) -- (3.4, 1.35) node [pos=0.5, anchor=north] {$b / 2$};
		\draw [ <-> ] (3.65, 1.55) -- (5.0, 2.9) node [pos=0.5, anchor=west] {$h$};
		\draw [ <-> ] (1.65, 1.53) -- (3.85, 1.97) node [pos=0.2, distance=-0.1., anchor=south] {$R$};
		
	\end{tikzpicture}
 \caption{Geometry of the three-point function on the lattice.} \label{fig:new2} \end{figure}
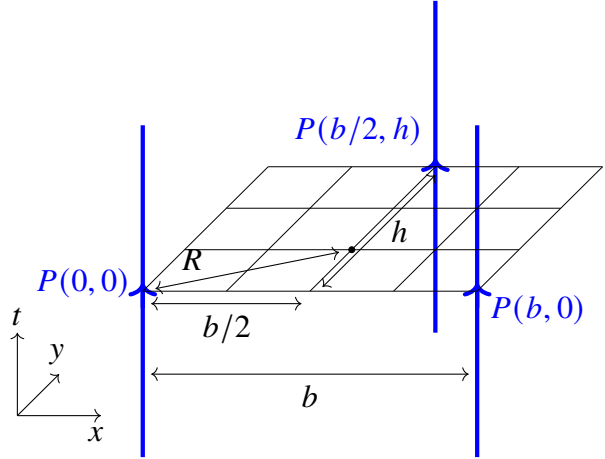

Using the same numerical setup of ref.~\cite{Caselle:2024zoh}, we study the system in the confining phase, both in the vicinity of the deconfinement transition ($0.75<T/T_c<0.9$) and in the low temperature regime ($T/T_c\simeq0.18$). At a fixed value of $\beta$ we investigate the three-point correlation function of Polyakov loops $G^{(3)}(x_1,x_2,x_3)$ as a function of their relative distance, using isosceles triangles which best approximate the equilateral geometry (see fig.~\ref{fig:new2}) and we change the temperature varying $N_t$. 
\subsection{Low-Temperature results}
Looking at the $1/R^2$ correction of the ground state energy $E_0(R)$ we estimated the Baryon Junciton Mass, performing the best fit of the $G^{(3)}(x_1,x_2,x_3)$ data, for each value of $\beta$ and $N_t$, with the functional form of eq.~\eqref{eqlowT}, where $A_{open}$, $\sigma$ and $M$ are the fit parameters. 

We report the results of the fits in tab.~\ref{tab:fit_res_baryon}.

\begin{table}[]
\centering
\resizebox{\textwidth}{!}{%
\begin{tabular}{|l|l|l|l|l|l|l||l|l|}
\hline
$\beta$ & $R_{min}/a$ & $R_{max}/a$ & $A_3\times10^{6}$          & $\sigma \, a^2$ & $aM$         & $\chi^2/N_{d.o.f}$ & $\sigma_0a^2$ from ref.~\cite{Caselle:2024zoh} &$M/\sqrt{\sigma}$ \\ \hline
$36.33$ & 4.64        & 12.68       & 7.85(58) & 0.009022(53) & 0.0137(13)  & 1.19  &  0.009344(17)  & 0.142(13)  \\ \hline
$39.65$ & 4.64        & 13.93       & 4.75(18) & 0.007518(23) & 0.01210(53) & 1.38  &  0.007831(58)  & 0.137(6)   \\ \hline
$42.97$ & 4.64        & 13.93       & 3.11(11) & 0.006378(18) & 0.01052(38) & 0.98  &  0.006645(59)  & 0.129(5)   \\ \hline
$46.29$ & 5.87        & 15.17       & 2.00(14) & 0.005460(27) & 0.01007(91) & 0.97  &  0.005716(59)  & 0.133(12)  \\ \hline
\end{tabular}
}  
\caption{Result of the best fit of the Polyakov-loop three-point correlator $G^{(3)}(R)$ for different values of $\beta$.}
\label{tab:fit_res_baryon}
\end{table}

Given that the ratio $M/\sqrt{\sigma}$ does not display any appreciable dependence on $\beta$, we carried out a combined fit over the four $\beta$ values, imposing a single common value of $M/\sqrt{\sigma}$ across all data sets.

\begin{figure}[!htb]
    \centering
    \includegraphics[width=\linewidth]{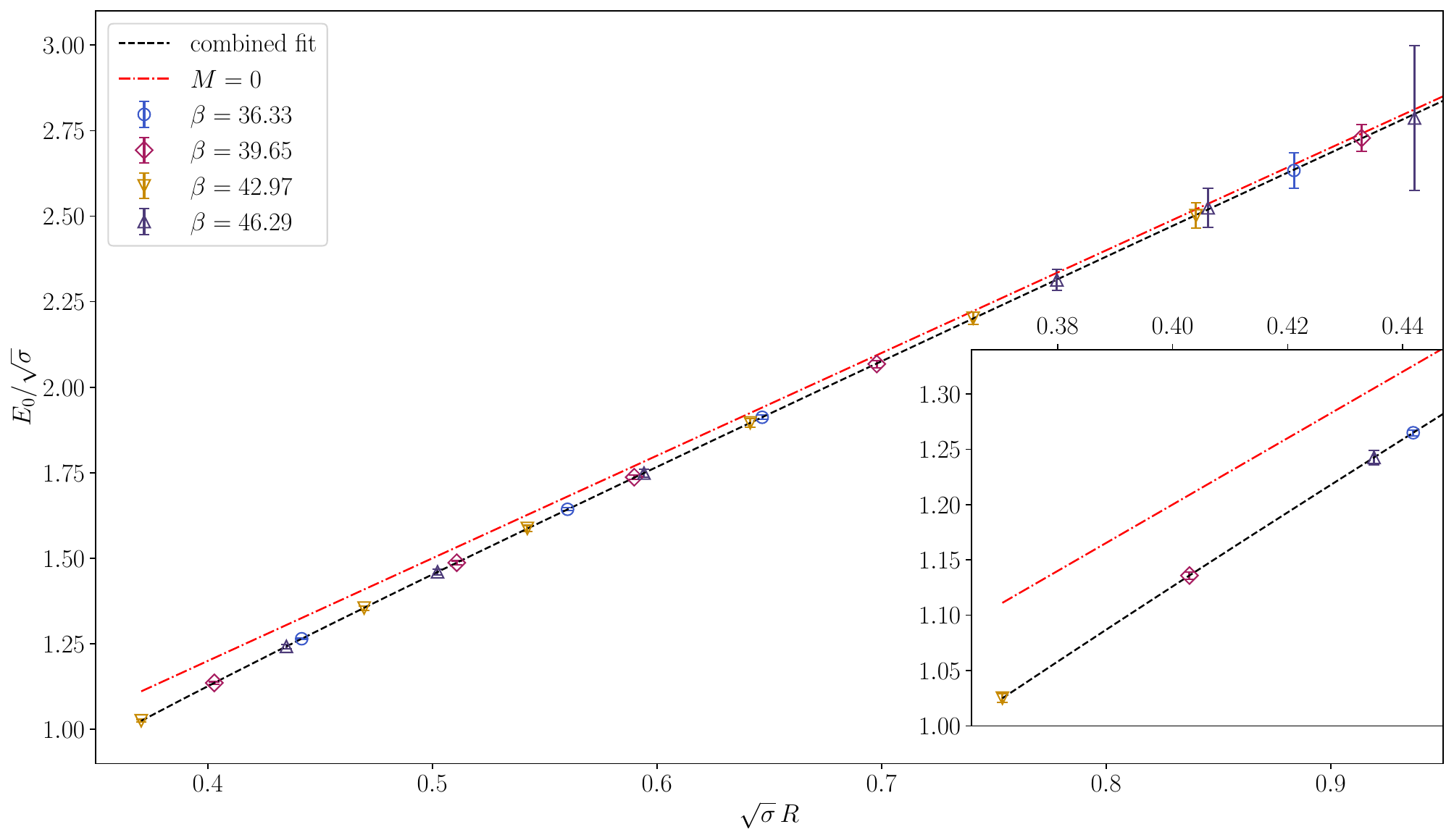}
\caption{Global best fits of the data at the four considered values of $\beta$, obtained using the fit ansatz of eq.~\eqref{eqlowT} and imposing a common value of $M/\sqrt{\sigma}$. No lattice-spacing dependence is observed for $E_0$. The parameters $A_3$ and $\sigma$ are determined from the combined fit, treating them as independent for each $\beta$. The red dash-dotted curve corresponds to the leading-order ground-state contribution with $M=0$, \textit{i.e.} $E_0 = 3 \sigma, R$. The inset focuses on the small-distance regime, where the inclusion of the $M$-dependent term is essential for an accurate data description.}
    \label{fig:global_fit}
\end{figure}

The global fit is displayed in fig.~\ref{fig:global_fit}. The corresponding result, which we take as our final determination of the baryon junction mass, reads as follows:
\begin{equation}
    \frac{M}{\sqrt{\sigma}}= 0.1355(36) ,\;\;\;\; \chi^2/N_{d.o.f}=1.04.
\end{equation}

Beyond its phenomenological interest ,this result, and especially its sign, indicate that the EST describing this configuration lies in the weakly coupled, perturbatively stable and in an unitary regime~\cite{Komargodski:2024swh}. Furthermore, the extracted value provides a nontrivial benchmark for models based on the AdS/QCD correspondence that aim at a non-perturbative description of confinement (see, for instance, \cite{Andreev:2021bfg}).

A significant prediction of the EST model is the equivalence of the string tension derived from the two-point function $\langle P^+(0)P(R) \rangle$ and that obtained from the baryonic state. As reported in tab.~\ref{tab:fit_res_baryon}, our analysis reveals a small but statistically significant mismatch between these two determinations. In particular, the string tension $\sigma$ resulting from the best fit is systematically lower by approximately $3$–$4\%$ compared to the value derived from the $\braket{P^+(0)P(R)}$ correlator (see the eighth column of tab.~\ref{tab:fit_res_baryon}). A similar deviation has recently been observed in high-precision studies of baryonic states in $(3+1)$ dimensions~\cite{Ma:2022vqf}. Although the difference is numerically modest, it becomes relevant in light of the precision achieved by the present data.
\subsection{High-Temperature results}
We used our high temperature results to test the Svetitsky--Yaffe mapping predictions. In particular, we fitted the data in the large distance regime with:
\begin{equation}\label{eq:long-distance}
G^{(3)}(R_Y) = A_3 \, K_0(R_Y E_0),
\end{equation} 
using $A_3$ and $E_0$ as free parameters of the fit.

From the large-distance determinations of $E_0$ we extract the string tension by adopting, for $E_0$, either the first-order EST expression of eq.~\eqref{eq5} or the functional form motivated by the Svetitsky--Yaffe correspondence, which indicates that the fully resummed EST result should be given by eq.~\eqref{eq8bis}. The string tension obtained using the first-order EST prediction is inconsistent with the values derived from the low-temperature fit reported in tab.~\ref{tab:fit_res_baryon}. By contrast, the determination based on the Svetitsky--Yaffe expression is compatible with the reference values extracted in the open-string channel.

\begin{figure}[htb]
    \centering
    \includegraphics[width=\linewidth]{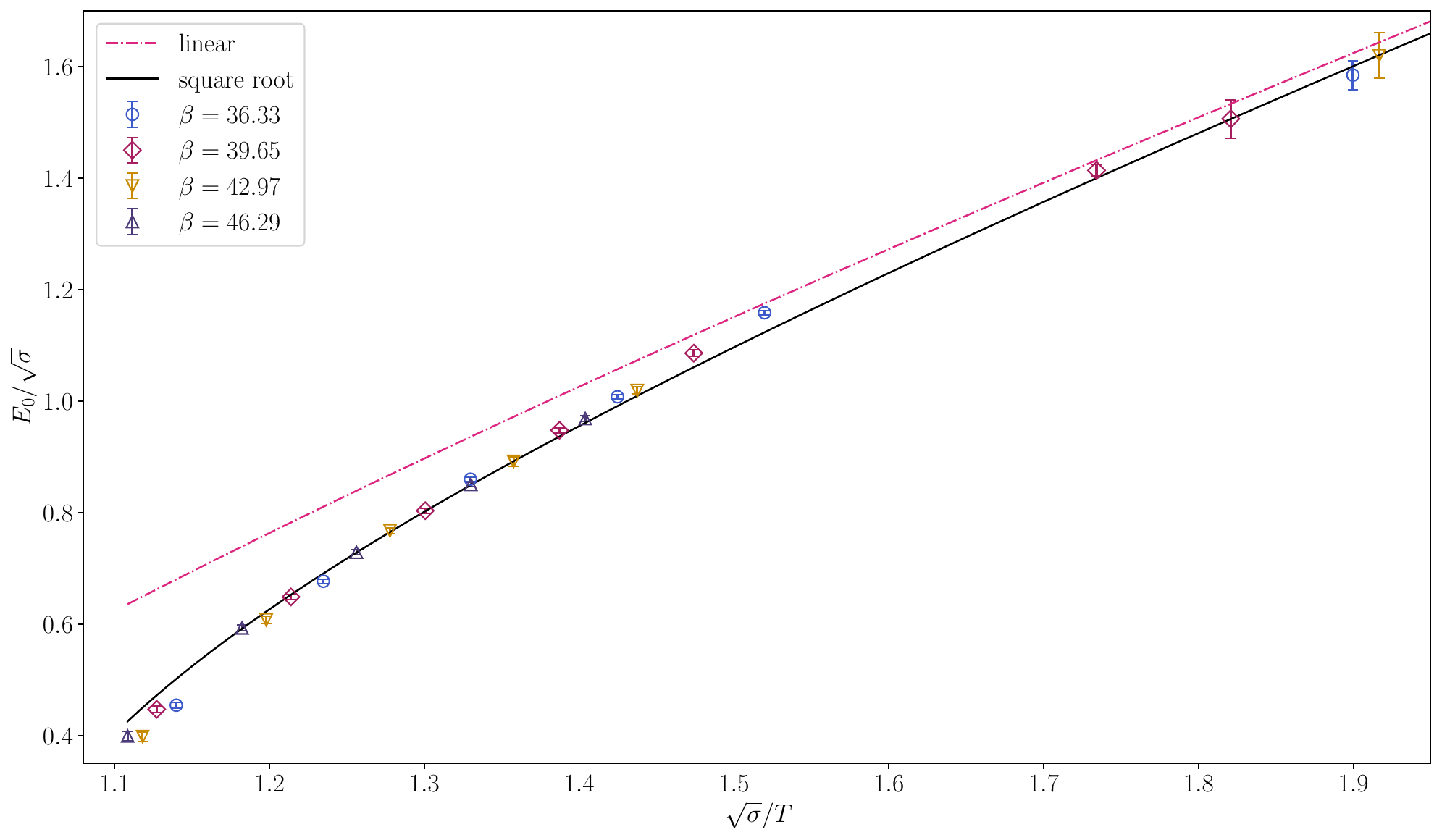}
    \caption{Estimates of $E_0$ in the high-temperature regime derived from large-distance fits using eq.~\eqref{eq:long-distance}. Both axes are scaled by low-temperature $\sigma$ values at the same lattice spacings. In these units, the models described by eq.~\eqref{eq5} and eq.~\eqref{eq9} have no free parameters, and are indicated by the red dash-dotted and black dashed lines, respectively. The data show good agreement with the latter, this suggest the string tension remains consistent between the closed (high-temperature) and open (low-temperature) channels, and that the square-root functional form provides a finer approximation of the ground state energy.}
    \label{fig:high_T_scaled}
\end{figure}

To further prove that the intuition suggested by the Svetitsky--Yaffe conjecture is well-founded, Fig.~\ref{fig:high_T_scaled} shows the high-temperature $E_0$ values scaled by $\sqrt{\sigma}$ (using low-temperature values). The agreement with the square root formula in eq.~\eqref{eq9} is evident, while the linear ansatz from eq.~\eqref{eq5} exhibits substantial deviations as the phase transition is approached.

Moreover, high-temperature $\sigma$ estimates align with low-temperature results but diverge from the values extracted from the two-point function in Ref.~\cite{Caselle:2024zoh}. This persistent discrepancy, also observed in the open string channel, indicates a systematic characteristic of the three-point function string tension. Such behavior may be linked to the baryon junction, possibly arising from flux tube broadening in its vicinity as described in Ref.~\cite{Pfeuffer:2008mz}.

\section{Conclusions and outlook}
This contribution presents a lattice numerical investigation of the three-string baryonic configuration in the $(2+1)$-dimensional $SU(3)$ Yang–Mills theory. Utilizing next-to-leading-order Effective String Theory (EST) expressions \cite{Komargodski:2024swh} and high-precision three-point Polyakov loop correlator data, the baryon junction mass $M$ was determined for the first time. This parameter is crucial for both theoretical and phenomenological developments. 

The obtained estimate of $M$ aligns with the phenomenological value used for the description of exotic hadrons \cite{Karliner:2016zzc}. However, since we studied the $(2+1)$-dimensional model, this  correspondence remains potentially coincidental. The sign and magnitude of $M$ indicate that the EST governing this configuration is situated in a stable, unitary, and weakly coupled regime \cite{Komargodski:2024swh}. Moreover, this result can be used to test non-perturbative confinement models derived from AdS/QCD duality.

A secondary result involves the high-temperature regime, just below the deconfinement temperature. Comparison with a three-state Potts model in one lower dimension yields excellent agreement, validating the Svetitsky–Yaffe mapping. This extends the findings for two-point functions \cite{Caselle:2024zoh} to the three-point sector. 

A very recent work \cite{Lou:2026xqr} extends the results from \cite{Komargodski:2024swh} to further perturbative orders. It will be interesting to test these novel findings with lattice numerical simulations.

Future investigations must address flux tube broadening near the baryon junction \cite{Pfeuffer:2008mz} using methods from \cite{Caselle:2026coc} to provide an independent validation of the EST framework. Additionally, extending this analysis to $(3+1)$-dimensional $SU(3)$ and full QCD is necessary to verify the universality of the $M/\sqrt{\sigma}$ ratio. Current evidence indicates that EST properties are preserved when coupled with matter \cite{Bonati:2021vbc}.

\vskip 1.5cm
\noindent {\large {\bf Acknowledgments}}
\vskip 0.2cm
We thank A. Bulgarelli, T. Canneti, E. Cellini, A. Mariani, A. Nada, and M. Panero for several useful discussions. 

We acknowledge support from the SFT Scientific Initiative of INFN. This work was partially supported by the Simons Foundation grant 994300 (Simons Collaboration on Confinement and QCD Strings) and by the Prin 2022 grant 2022ZTPK4E. The numerical simulations have been performed on the Leonardo machine at CINECA, based on the agreement with INFN, under the project INF25\_sft.

\bibliographystyle{JHEP}
\bibliography{biblio}

@article{Caselle:2025elf,
    author = "Caselle, Michele and Magnoli, Nicodemo and Panfalone, Dario and Verzichelli, Lorenzo",
    title = "{The mass of the baryon junction: a lattice computation in 2 + 1 dimensions}",
    eprint = "2508.00608",
    archivePrefix = "arXiv",
    primaryClass = "hep-lat",
    doi = "10.1007/JHEP12(2025)019",
    journal = "JHEP",
    volume = "12",
    pages = "019",
    year = "2025"
}

@article{Caselle:2026coc,
    author = "Caselle, Michele and Cellini, Elia and Nada, Alessandro and Panfalone, Dario and Verzichelli, Lorenzo",
    title = "{Intrinsic Width of the Flux Tube as a tool to explore confining mechanisms in Lattice Gauge Theories}",
    eprint = "2601.19520",
    archivePrefix = "arXiv",
    primaryClass = "hep-lat",
    month = "1",
    year = "2026"
}

@Article{ Nambu:1974zg,
	author = "Yoichiro Nambu",
	title = "{Strings, Monopoles and Gauge Fields}",
	journal = "Phys. Rev.",
	volume = "D10",
	year = "1974",
	pages = "4262",
	doi = "10.1103/PhysRevD.10.4262",
	reportNumber = "EFI 74/40",
	SLACcitation = "%%CITATION = PHRVA,D10,4262;%%"
}

@Article{ Goto:1971ce,
	Title = "{Relativistic quantum mechanics of one-dimensional mechanical continuum and subsidiary condition of dual resonance model}",
	Author = "Tetsuo Goto",
	Journal = "Prog.Theor.Phys.",
	Year = "1971",
	Pages = "1560--1569",
	Volume = "46",
	Doi = "10.1143/PTP.46.1560"
}

@Article{ Luscher:1980ac,
	Title = "{Symmetry Breaking Aspects of the Roughening Transition in Gauge Theories}",
	Author = "M. Luscher",
	Journal = "Nucl.Phys.",
	Year = "1981",
	Pages = "317",
	Volume = "B180",
	Doi = "10.1016/0550-3213(81)90423-5"
}

@Article{ Luscher:2004ib,
	Title = "{String excitation energies in SU(N) gauge theories beyond the free-string approximation}",
	Author = "Martin Luscher and Peter Weisz",
	Journal = "JHEP",
	Year = "2004",
	Pages = "014",
	Volume = "07",
	Archiveprefix = "arXiv",
	Doi = "10.1088/1126-6708/2004/07/014",
	Eprint = "hep-th/0406205",
	Slaccitation = "%%CITATION = HEP-TH/0406205;%%"
}

@Article{ Polchinski:1991ax,
	Title = "{Effective string theory}",
	Author = "Joseph Polchinski and Andrew Strominger",
	Journal = "Phys.Rev.Lett.",
	Year = "1991",
	Pages = "1681--1684",
	Volume = "67",
	Doi = "10.1103/PhysRevLett.67.1681"
}

@article{Brandt:2016xsp,
      author         = "Brandt, Bastian B. and Meineri, Marco",
      title          = "{Effective string description of confining flux tubes}",
      year           = "2016",
      journal        = "Int. J. Mod. Phys.",
      volume         = "A31",
      pages          = "1643001",
      eprint         = "1603.06969",
      archivePrefix  = "arXiv",
      primaryClass   = "hep-th",
      SLACcitation   = "%%CITATION = ARXIV:1603.06969;%%",
}

@Article{ Aharony:2013ipa,
	Title = "{The Effective Theory of Long Strings}",
	Author = "Ofer Aharony and Zohar Komargodski",
	Journal = "JHEP",
	Year = "2013",
	Pages = "118",
	Volume = "1305",
	Archiveprefix = "arXiv",
	Doi = "10.1007/JHEP05(2013)118",
	Eprint = "1302.6257",
	Primaryclass = "hep-th",
	Reportnumber = "WIS-01-13-FEB-DPPA",
	Slaccitation = "%%CITATION = ARXIV:1302.6257;%%"
}

@article{Caselle:2021eir,
    author = "Caselle, Michele",
    title = "{Effective String Description of the Confining Flux Tube at Finite Temperature}",
    eprint = "2104.10486",
    archivePrefix = "arXiv",
    primaryClass = "hep-lat",
    doi = "10.3390/universe7060170",
    journal = "Universe",
    volume = "7",
    number = "6",
    pages = "170",
    year = "2021"
}

@Article{ Billo:2012da,
	author = "M. Billo and M. Caselle and F. Gliozzi and M. Meineri and R. Pellegrini",
	title = "{The Lorentz-invariant boundary action of the confining string and its universal contribution to the inter-quark potential}",
	journal = "JHEP",
	volume = "05",
	year = "2012",
	pages = "130",
	doi = "10.1007/JHEP05(2012)130",
	eprint = "1202.1984",
	archivePrefix = "arXiv",
	primaryClass = "hep-th",
	reportNumber = "DFTT-02-2012",
	SLACcitation = "%%CITATION = ARXIV:1202.1984;%%"
}

@article{Gliozzi:2012cx,
      author         = "Gliozzi, Ferdinando and Meineri, Marco",
      title          = "{Lorentz completion of effective string (and p-brane)
                        action}",
      journal        = "JHEP",
      volume         = "1208",
      pages          = "056",
      doi            = "10.1007/JHEP08(2012)056",
      year           = "2012",
      eprint         = "1207.2912",
      archivePrefix  = "arXiv",
      primaryClass   = "hep-th",
      SLACcitation   = "%%CITATION = ARXIV:1207.2912;%%",
}

@article{Caselle:2024zoh,
    author = "Caselle, Michele and Magnoli, Nicodemo and Nada, Alessandro and Panero, Marco and Panfalone, Dario and Verzichelli, Lorenzo",
    title = "{Confining strings in three-dimensional gauge theories beyond the Nambu-Got\={o} approximation}",
    eprint = "2407.10678",
    archivePrefix = "arXiv",
    primaryClass = "hep-lat",
    doi = "10.1007/JHEP08(2024)198",
    journal = "JHEP",
    volume = "08",
    pages = "198",
    year = "2024"
}

@article{Pfeuffer:2008mz,
    author = "Pfeuffer, Melanie and Bali, Gunnar S. and Panero, Marco",
    title = "{Fluctuations of the baryonic flux-tube junction from effective string theory}",
    eprint = "0810.1649",
    archivePrefix = "arXiv",
    primaryClass = "hep-th",
    doi = "10.1103/PhysRevD.79.025022",
    journal = "Phys. Rev. D",
    volume = "79",
    pages = "025022",
    year = "2009"
}

@article{Komargodski:2024swh,
    author = "Komargodski, Zohar and Zhong, Siwei",
    title = "{Baryon junction and string interactions}",
    eprint = "2405.12005",
    archivePrefix = "arXiv",
    primaryClass = "hep-th",
    doi = "10.1103/PhysRevD.110.056018",
    journal = "Phys. Rev. D",
    volume = "110",
    number = "5",
    pages = "056018",
    year = "2024"
}

@article{Bakry:2014gea,
    author = "Bakry, Ahmed S. and Chen, Xurong and Zhang, Peng-Ming",
    title = "{Y-stringlike behavior of a static baryon at finite temperature}",
    eprint = "1412.3568",
    archivePrefix = "arXiv",
    primaryClass = "hep-lat",
    doi = "10.1103/PhysRevD.91.114506",
    journal = "Phys. Rev. D",
    volume = "91",
    pages = "114506",
    year = "2015"
}

@article{Borisenko:2018zzd,
    author = "Borisenko, O. and Chelnokov, V. and Mendicelli, Emanuele and Papa, Alessandro",
    title = "{Three-quark potentials in an $SU(3)$ effective Polyakov loop model}",
    eprint = "1812.05384",
    archivePrefix = "arXiv",
    primaryClass = "hep-lat",
    doi = "10.1016/j.nuclphysb.2019.02.002",
    journal = "Nucl. Phys. B",
    volume = "940",
    pages = "214--238",
    year = "2019"
}

@article{Koma:2017hcm,
    author = "Koma, Yoshiaki and Koma, Miho",
    title = "{Precise determination of the three-quark potential in SU(3) lattice gauge theory}",
    eprint = "1703.06247",
    archivePrefix = "arXiv",
    primaryClass = "hep-lat",
    doi = "10.1103/PhysRevD.95.094513",
    journal = "Phys. Rev. D",
    volume = "95",
    number = "9",
    pages = "094513",
    year = "2017"
}

@article{Andreev:2021bfg,
    author = "Andreev, Oleg",
    title = "{Remarks on static three-quark potentials, string breaking and gauge/string duality}",
    eprint = "2101.03858",
    archivePrefix = "arXiv",
    primaryClass = "hep-ph",
    reportNumber = "LMU-ASC 47/20",
    doi = "10.1103/PhysRevD.104.026005",
    journal = "Phys. Rev. D",
    volume = "104",
    number = "2",
    pages = "026005",
    year = "2021"
}

@article{Ma:2022vqf,
    author = "Ma, Yao and Meng, Lu and Chen, Yan-Ke and Zhu, Shi-Lin",
    title = "{Ground state baryons in the flux-tube three-body confinement model using diffusion Monte~Carlo}",
    eprint = "2211.09021",
    archivePrefix = "arXiv",
    primaryClass = "hep-ph",
    doi = "10.1103/PhysRevD.107.054035",
    journal = "Phys. Rev. D",
    volume = "107",
    number = "5",
    pages = "054035",
    year = "2023"
}

@article{Jahn:2003uz,
    author = "Jahn, Oliver and de Forcrand, Philippe",
    editor = "Aoki, S. and Hashimoto, S. and Ishizuka, N. and Kanaya, K. and Kuramashi, Y.",
    title = "{Baryons and confining strings}",
    eprint = "hep-lat/0309115",
    archivePrefix = "arXiv",
    reportNumber = "CERN-TH-2003-217",
    doi = "10.1016/S0920-5632(03)02685-9",
    journal = "Nucl. Phys. B Proc. Suppl.",
    volume = "129",
    pages = "700--702",
    year = "2004"
}

@article{deForcrand:2005vv,
    author = "de Forcrand, Ph. and Jahn, Oliver",
    editor = "Guidal, M. and Kunne, F. and Soyeur, M. and Pire, B.",
    title = "{The Baryon static potential from lattice QCD}",
    eprint = "hep-ph/0502039",
    archivePrefix = "arXiv",
    reportNumber = "CERN-PH-TH-2005-020",
    doi = "10.1016/j.nuclphysa.2005.03.127",
    journal = "Nucl. Phys. A",
    volume = "755",
    pages = "475--480",
    year = "2005"
}

@article{Dmitrasinovic:2009ma,
    author = "Dmitrasinovic, V. and Sato, Toru and Suvakov, Milovan",
    title = "{On the smooth cross-over transition from the Delta-string to the Y-string three-quark potential}",
    eprint = "0908.2687",
    archivePrefix = "arXiv",
    primaryClass = "hep-ph",
    doi = "10.1103/PhysRevD.80.054501",
    journal = "Phys. Rev. D",
    volume = "80",
    pages = "054501",
    year = "2009"
}

@article{Cardoso:2008sb,
    author = "Cardoso, M. and Bicudo, P.",
    title = "{First study of the three-gluon static potential in Lattice QCD}",
    eprint = "0807.1621",
    archivePrefix = "arXiv",
    primaryClass = "hep-lat",
    doi = "10.1103/PhysRevD.78.074508",
    journal = "Phys. Rev. D",
    volume = "78",
    pages = "074508",
    year = "2008"
}

@article{Huebner:2007yzb,
    author = "Huebner, Kay and Karsch, Frithjof and Kaczmarek, Olaf and Vogt, Oliver",
    title = "{Heavy quark free energies for three quark systems at finite temperature}",
    eprint = "0710.5147",
    archivePrefix = "arXiv",
    primaryClass = "hep-lat",
    reportNumber = "BNL-NT-07-44, BI-TP-2007-31",
    doi = "10.1103/PhysRevD.77.074504",
    journal = "Phys. Rev. D",
    volume = "77",
    pages = "074504",
    year = "2008"
}

@article{Brambilla:2009cd,
    author = "Brambilla, Nora and Ghiglieri, Jacopo and Vairo, Antonio",
    title = "{The Three-quark static potential in perturbation theory}",
    eprint = "0911.3541",
    archivePrefix = "arXiv",
    primaryClass = "hep-ph",
    reportNumber = "IFUM-915-FT, TUM-EFT-1-09, IFUM-915-FT TUM-EFT 1/09",
    doi = "10.1103/PhysRevD.81.054031",
    journal = "Phys. Rev. D",
    volume = "81",
    pages = "054031",
    year = "2010",
    note = "[Erratum: Phys.Rev.D 107, 019904 (2023)]"
}

@article{Caselle:2005sf,
    author = "Caselle, M. and Delfino, G. and Grinza, P. and Jahn, O. and Magnoli, N.",
    title = "{Potts correlators and the static three-quark potential}",
    eprint = "hep-th/0511168",
    archivePrefix = "arXiv",
    reportNumber = "DFTT-29-05, SISSA-87-2005-FM, PTA-05-72, MIT-CPT-3705, GEF-TH-9-05",
    doi = "10.1088/1742-5468/2006/03/P03008",
    journal = "J. Stat. Mech.",
    volume = "0603",
    pages = "P03008",
    year = "2006"
}

@article{Sakumichi:2015rfa,
    author = "Sakumichi, Naoyuki and Suganuma, Hideo",
    title = "{Three-quark potential and Abelian dominance of confinement in SU(3) QCD}",
    eprint = "1501.07596",
    archivePrefix = "arXiv",
    primaryClass = "hep-lat",
    doi = "10.1103/PhysRevD.92.034511",
    journal = "Phys. Rev. D",
    volume = "92",
    number = "3",
    pages = "034511",
    year = "2015"
}

@article{Alexandrou:2002sn,
    author = "Alexandrou, C. and de Forcrand, P. and Jahn, Oliver",
    editor = "Edwards, R. and Negele, John W. and Richards, D.",
    title = "{The Ground state of three quarks}",
    eprint = "hep-lat/0209062",
    archivePrefix = "arXiv",
    doi = "10.1016/S0920-5632(03)01659-1",
    journal = "Nucl. Phys. B Proc. Suppl.",
    volume = "119",
    pages = "667--669",
    year = "2003"
}

@article{Takahashi:2002bw,
    author = "Takahashi, Toru T. and Suganuma, H. and Nemoto, Y. and Matsufuru, H.",
    title = "{Detailed analysis of the three quark potential in SU(3) lattice QCD}",
    eprint = "hep-lat/0204011",
    archivePrefix = "arXiv",
    doi = "10.1103/PhysRevD.65.114509",
    journal = "Phys. Rev. D",
    volume = "65",
    pages = "114509",
    year = "2002"
}

@article{Bissey:2006bz,
    author = "Bissey, F. and Cao, F-G. and Kitson, A. R. and Signal, A. I. and Leinweber, D. B. and Lasscock, B. G. and Williams, A. G.",
    title = "{Gluon flux-tube distribution and linear confinement in baryons}",
    eprint = "hep-lat/0606016",
    archivePrefix = "arXiv",
    reportNumber = "ADP-06-04-T635",
    doi = "10.1103/PhysRevD.76.114512",
    journal = "Phys. Rev. D",
    volume = "76",
    pages = "114512",
    year = "2007"
}

@article{Guida:1995kc,
    author = "Guida, Riccardo and Magnoli, Nicodemo",
    title = "{All order IR finite expansion for short distance behavior of massless theories perturbed by a relevant operator}",
    eprint = "hep-th/9511209",
    archivePrefix = "arXiv",
    reportNumber = "GEF-TH-12-1995",
    doi = "10.1016/0550-3213(96)00175-7",
    journal = "Nucl. Phys. B",
    volume = "471",
    pages = "361--388",
    year = "1996"
}

@article{Bonati:2021vbc,
    author = "Bonati, Claudio and Caselle, Michele and Morlacchi, Silvia",
    title = "{The Unreasonable effectiveness of effective string theory: The case of the 3D SU(2) Higgs model}",
    eprint = "2106.08784",
    archivePrefix = "arXiv",
    primaryClass = "hep-lat",
    doi = "10.1103/PhysRevD.104.054501",
    journal = "Phys. Rev. D",
    volume = "104",
    number = "5",
    pages = "054501",
    year = "2021"
}

@article{Karliner:2016zzc,
    author = "Karliner, Marek and Nussinov, Shmuel and Rosner, Jonathan L.",
    title = "{$Q Q \bar Q \bar Q$ states: masses, production, and decays}",
    eprint = "1611.00348",
    archivePrefix = "arXiv",
    primaryClass = "hep-ph",
    reportNumber = "EFI-16-22, TAUP-3012-16",
    doi = "10.1103/PhysRevD.95.034011",
    journal = "Phys. Rev. D",
    volume = "95",
    number = "3",
    pages = "034011",
    year = "2017"
}

@article{Lou:2026xqr,
    author = "Lou, Xuzixiang and Zhong, Siwei",
    title = "{Baryon Junction and String Interactions: Part II}",
    eprint = "2602.17771",
    archivePrefix = "arXiv",
    primaryClass = "hep-th",
    month = "2",
    year = "2026"
}
\end{document}